\documentstyle[preprint,aps,epsf]{revtex}
\draft
\begin{document}
\preprint{Phys. Rev. C (1998) in press.}
\title{Isospin relaxation time in heavy-ion 
collisions at intermediate energies}
\bigskip
\author{\bf Bao-An Li and C.M. Ko}
\address{Cyclotron Institute and Department of Physics\\ 
Texas A\&M University, College Station, TX 77843, USA}
\maketitle

\begin{quote}
Using an isospin-dependent transport model, we have studied the 
isospin and momentum relaxation times in the heavy residues formed 
in heavy-ion collisions at intermediate energies. It is found that only 
at incident energies below the Fermi energy, chemical or thermal 
equilibrium can be reached before dynamical instability is developed
in the heavy residues.
Also, the isospin relaxation time is shorter (longer) than 
that for momentum at beam energies lower (higher) than the Fermi energy. 

\end{quote}

\newpage

In studying heavy-ion collisions at intermediate energies, it is important
to know if thermodynamical, i.e., thermal (momentum) and chemical (isospin), 
equilibrium can be reached in the heavy residues formed in these collisions.
Although the early stage of heavy ion collisions 
must be described by microscopic models, a macroscopic treatment of 
the later stage in terms of temperature, volume and 
chemical potential becomes possible if thermodynamical equilibrium is
established.  Also, the interpretation of nuclear multifragmentation in 
heavy ion collisions either as a dynamical or as a statistical process
depends on whether thermodynamical equilibrium is achieved in the collisions.
Based on the assumption that thermodynamical equilibrium can be established 
in heavy-ion collisions, many statistical models have been developed, and 
they seem to be quite successful in describing the experimental data.
A critical examination of thermodynamical equilibrium in heavy ion 
collisions is thus needed. A review on this subject can be found 
in Ref. \cite{moretto}. 

In past few years, extensive theoretical efforts have been devoted to 
investigate the rate of thermalization of nuclear matter 
in intermediate energy heavy ion collisions.  Using the
Uehling-Uhlenbeck equation, momentum relaxation both in
collisions of semi-infinite nuclear matter and in infinite nuclear 
matter have been studied \cite{bertsch78,randrup79,wong82}. 
Collisions of finite nuclei have also been studied using the 
Boltzmann-Uehling-Uhlenbeck equation \cite{cas87,abg94,had96,bor97}. 
On the other hand, only limited studies have been carried out for 
understanding the rate of chemical equilibrium in heavy ion 
collisions \cite{bali95,don97}. In both statistical and 
dynamical models it is usually assumed that chemical equilibrium is reached
either instantaneously or as fast as thermal equilibrium. This is supported
by deep inelastic heavy ion collisions at low energies, where
the isospin degree of freedom has been found to first reach 
equilibrium \cite{gatty75,fed78,udo84}. 
Recent experimental studies \cite{sherry1,sherry2,sherry3,sherry4}
of the isotopic composition of intermediate 
mass fragments (IMF) and their angular distributions in heavy ion 
collisions have also shown that only at low energies ($\approx 30$ MeV/nucleon)
isospin equilibrium is reached before their emission.

With recent advances in experiments using
radioactive beams with large neutron or proton excess, it has been
possible to create a transient state of nuclear matter with appreciable 
isospin asymmetry.  Knowledge on the rate of chemical and thermal 
equilibrium in such an asymmetric nuclear matter is thus needed \cite{bali97}. 

In this Brief Report, we shall report the results from a study of comparing 
the rate of chemical equilibrium to that of thermal equilibrium
using an isospin-dependent Boltzmann-Uehling-Uhlenbeck (BUU) transport model. 
The model has been quite successful in predicting and 
explaining a number of isospin-dependent phenomena in 
heavy-ion collisions at intermediate
energies \cite{bali95,bali96,pak1,pak2,lik97}.
The isospin dependence is introduced in the model through using
different total and differential cross sections, Pauli blocking and mean fields for protons
and neutrons. A recent review of this model, especially its
isospin-dependent aspects, is given in Ref. \cite{bali97}. 

In the model study of heavy ion collisions, a heavy residue can be identified
as a collection of nucleons with densities higher than 1/10 of normal
nuclear matter density. 
To characterize the degree of chemical equilibrium, we introduce the 
following quantity
\begin{equation}
\lambda_I(t)\equiv \frac{(n/p)_{y>0}}{(n/p)_{y<0}},
\end{equation}
where $(n/p)_{y>0}$ and $(n/p)_{y<0}$ are, respectively, the neutron to 
proton ratio for positive and negative rapidity nucleons in the rest 
frame of the residue. If chemical equilibrium is established
this quantity then has a value of one.
We further define the isospin relaxation time $\tau_I$ as the time
when the quantity $(\lambda_I(t)-1)/(\lambda_I(0)-1)$ is 0.01, i.e., it is
one percent from its equilibrium value. This time is an approximate 
measure of the rate at which the residue reaches chemical equilibrium.
We note that this definition is somewhat different from that one would
usually use, i.e., $\lambda(\tau_I)=1/e$.

For describing thermal equilibrium of the heavy residue, we use 
the quadrupole moment $Q_{zz}(t)$ in its rest frame, i.e., 
\begin{equation}
Q_{zz}(t)=\int\frac{d\vec{r}d\vec{p}}{(2\pi)^3}(2p^2_z-p^2_x-p^2_y)
f(\vec{r},\vec{p},t),
\end{equation}
where $f(\vec{r},\vec{p},t)$ is the Wigner function from the BUU model 
calculations. Obviously $Q_{zz}=0$ is a necessary, although not a 
sufficient, condition for thermal equilibrium.
Similar to the definition of $\tau_I$ we define the momentum
relaxation time $\tau_p$ as the time when $Q_{zz}(t)/Q_{zz}(0)=1\%$.
This quantity then measures the rate at which the residue reaches thermal
equilibrium.

Another important property of the heavy residue is possible existence
of dynamical instability. To study this phenomenon, we introduce
the square of the adiabatic sound velocity \cite{bauer92,ligross}
\begin{equation}
  v_s^2 = \frac{1}{m}\left(\partial P\over\partial\rho\right)_S
  = {1\over m} \left[ {10\over 9}\langle E_k \rangle
	                      + a {\rho\over\rho_0}
			      + b \sigma\left(\rho\over\rho_0\right)^{\sigma}
			\right]\ ,
\end{equation}
where $\langle E_k \rangle$ is the average kinetic
energy per nucleon, $a=$ -358.1 MeV, $b=$ 304.8 MeV and $\sigma=7/6$ are 
the parameters corresponding to a soft nuclear equation of state.
For $v^2_s < 0$, a homogeneous nuclear matter is unstable against the 
growth of fluctuation, leading to dynamical instability or spinodal 
decomposition,  

We have studied the time dependence of
$(\lambda_I(t)-1)/(\lambda_I(0)-1)$ and $Q_{zz}(t)/Q_{zz}(0)$ compared to
that of $v_s^2(t)/v_s^2(0)$. They are shown in Fig. 1
for collisions of $^{40}$Ca+$^{124}$Sn at an impact
parameter of 1 fm and at beam energies of 25, 50, 150 and 300 MeV/nucleon.
The system considered here has an initial $\lambda_I(0)=1.48$.
For collisions at a beam energy of 25 MeV/nucleon the residue
is found to be dynamically stable up to 300 fm/c during the collision. 
This time interval is long enough for both thermal and
chemical equilibrium to be fully established as shown in the middle and
lower windows of the first column. On the other hand, for collisions at beam
energies above 50 MeV/nucleon a significant compression appears, and 
this is followed by expansion, leading into the adiabatic
spinodal region after about 50 fm/c. We notice that at this time the heavy 
residue formed in the collision is still far from thermal and chemical 
equilibrium. Both the momentum and isospin 
asymmetries of the heavy residue are seen to change oscillatorily with time. 
The study on the nature of this oscillation in isospin asymmetry, 
in particular its possible connection to the giant dipole resonance in 
nuclei, is underway and will be reported elsewhere.
 
We also notice from the middle window of the fourth column that nuclear
transparency occurs at E/A=300 MeV.
After spinodal decomposition the heavy residue quickly starts
to break up into fragments and nucleons \cite{ligross}. Although
the isotopic contents of these fragments and nucleons depend on 
the emission angle, it is also strongly influenced by 
the neutron to proton ratio of the target and projectile in the 
entrance channel instead the average neutron to proton ratio of the 
combined system. This observation is consistent with recent experimental 
findings \cite{sherry1,sherry2,sherry3,sherry4}.

Although chemical and thermal equilibrium are not completely established
at beam energies higher than the Fermi energy, it is still interesting to
compare their relaxation times.  This is shown in Fig. 2. In the left window,
the comparison is made for $^{40}$Ca+$^{124}$Sn collisions at beam energies 
from 25 to 300 MeV/nucleon and at an impact parameter of 1 fm.
The momentum relaxation time is found to decrease with increasing
beam energy. This is in qualitative agreement with that found in
Refs. \cite{bertsch78,randrup79,wong82,cas87,abg94,bor97}. On the other hand, 
the isospin relaxation time decreases slowly with the beam energy.
The shorter isospin relaxation time at incident energies below about 50
MeV/nucleon is in agreement with what was found in deep inelastic heavy ion 
collisions \cite{gatty75,fed78,udo84}. At higher incident energies the time
for reaching momentum equilibrium is found shorter than that for isospin 
equilibrium. For example, a 20 fm/c difference in the relaxation time 
is observed at E/A=300 MeV. In the right window, the isospin and 
momentum relaxation time as functions of the initial isospin asymmetry 
$\lambda_I(0)\equiv (n/p)_{\rm projectile}/(n/p)_{\rm target}$ are compared 
for $^{40}$Ca induced reactions on several isobaric targets of mass 124 at
a beam energy of 300 MeV and at an impact parameter of 1 fm. It is seen that 
although the momentum relaxation is almost independent of the initial isospin 
asymmetry, the isospin relaxation time increases with the initial isospin 
asymmetry. 

The short relaxation time for isospin than momentum at 
low incident energies can be understood as follows. First, nucleon-nucleon 
collisions, which are responsible for momentum relaxation, are more likely 
to be suppressed due to Pauli blocking. Secondly, the repulsive symmetry 
potential for neutrons and the attractive symmetry potential for protons 
make preequilibrium emissions of neutrons more likely than protons in low
energy collisions, as shown in Ref. \cite{lik97}, which thus enhances
the isospin relaxation rate in the residue. On the other hand,
in high energy collisions, Pauli blocking is less effective and the symmetry 
potential is also less important, leading thus to a shorter momentum 
relaxation time and a longer isospin relaxation time.

We have also studied effects due to different forms 
of symmetry potential and the charge exchange reaction ($pn\rightarrow np$) 
on chemical and thermal equilibrium.  They are found to have no discernible
effects on both the momentum and isospin relaxation times. Only during
the later stage of the collisions, they affect slightly the momentum and
isospin distributions.

In summary, using an isospin-dependent BUU model we have studied
the isospin and momentum relaxation times in the heavy residues formed in 
heavy-ion collisions at intermediate energies. 
They are compared both with each other and also with the time for 
dynamical instability to appear in the heavy residue.
We have found that chemical and thermal 
equilibrium can be completely established only at beam energies 
below the Fermi energy. At higher energies the dynamical 
instability sets in before either chemical or thermal 
equilibrium is achieved. Moreover, the isospin relaxation 
time is shorter (longer) than that for momentum at beam energies lower 
(higher) than the Fermi energy. 

This work was supported in part by the NSF Grant No. PHY-9509266 and 
the Robert A Welch Foundation under Grant A-1358.

\section*{Figure Captions}

\begin{description}

\item{ Fig. 1} \ \ \ 
The development of dynamical instability (upper windows), thermal (middle
windows) and chemical (lower windows) equilibrium in $^{40}$Ca+$^{124}$Sn
collisions at an impact parameter of 1 fm and beam energies of 25, 50, 150
and 300 MeV/nucleon, respectively.
\item{ Fig. 2} \ \ \
Left window: isospin (open circles) and momentum (filled circles)
relaxation times as functions of beam energy for the reactions
shown in Fig. 1. Right window: relaxation times as functions
of $\lambda_I(0)$ of the reactions. 
\end{description}

\begin{figure}[htp]
\vspace{-5.0truecm}
\setlength{\epsfxsize=10truecm}
\centerline{\epsffile{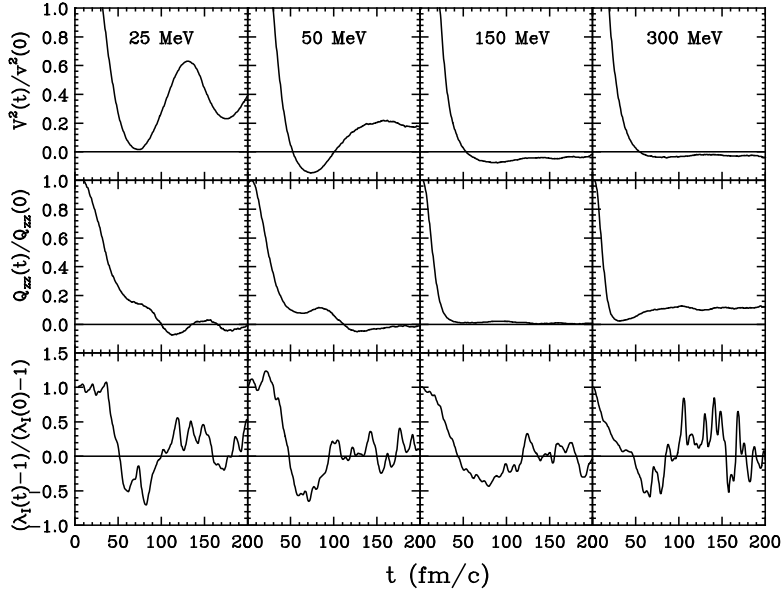} 
\caption{The development of dynamical instability (upper windows), thermal 
(middle windows) and chemical (lower windows) equilibrium in 
$^{40}$Ca+$^{124}$Sn
collisions at an impact parameter of 1 fm and beam energies of 25, 50, 150
and 300 MeV/nucleon, respectively.}}
\end{figure}  
\newpage
\begin{figure}[htp]
\vspace{-5.0truecm}
\setlength{\epsfxsize=10truecm}
\centerline{\epsffile{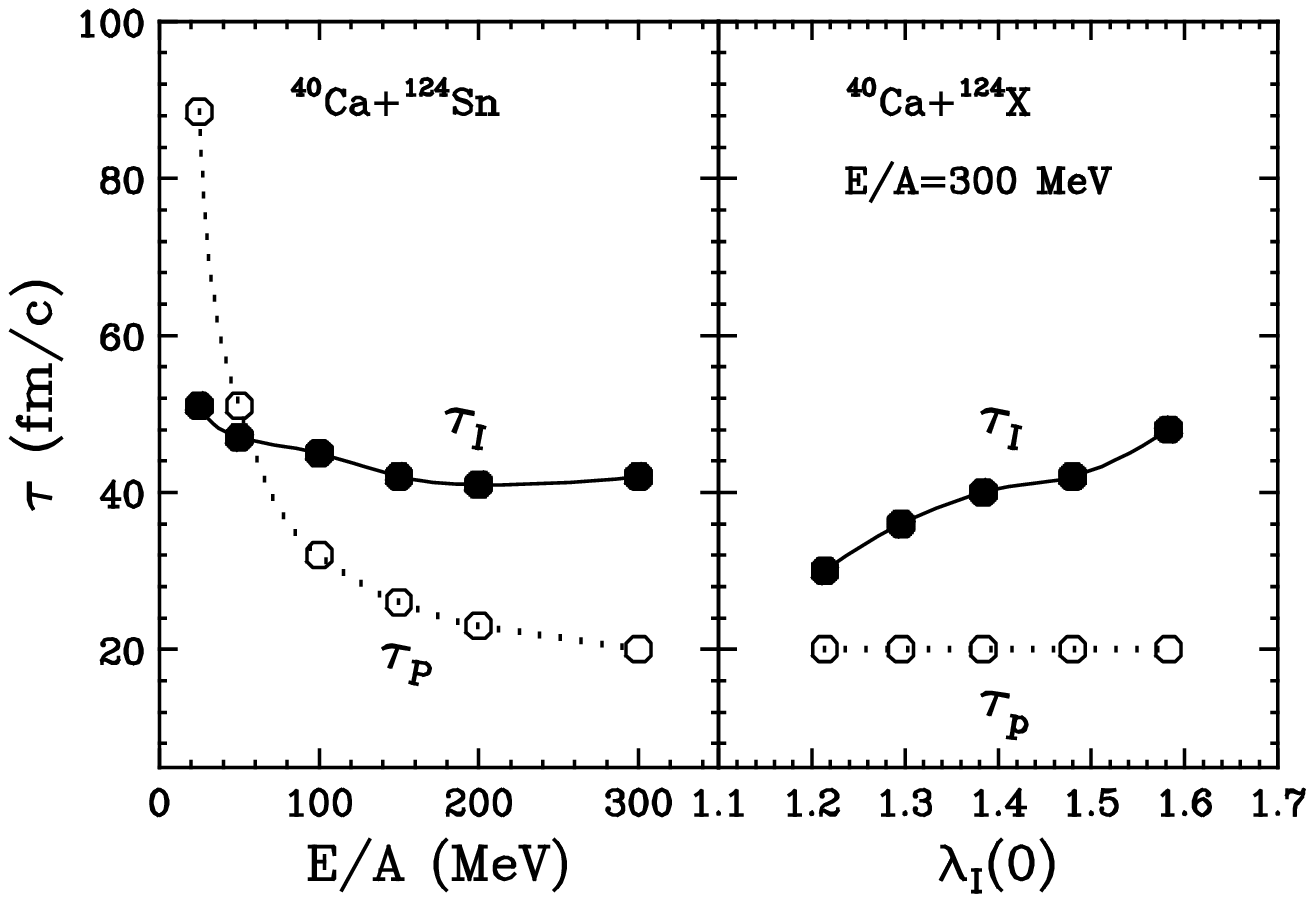} 
\caption{Left window: isospin (open circles) and momentum (filled circles)
relaxation times as functions of beam energy for the reactions
shown in Fig. 1. Right window: relaxation times as functions
of $\lambda_I(0)$ of the reactions.}}
\end{figure}  


\begin{thebibliography}{99}

\bibitem{moretto}L. G. Moretto and G. J. Wozniak, Ann. Rev. Nucl. Part. Sci. 
{\bf 43}, 123 (1993).
\vspace{-8pt}
\bibitem{bertsch78}G. F. Bertsch, Z. Phys. A {\bf 289}, 103 (1978).
\vspace{-8pt}
\bibitem{randrup79}J. Randrup, Nucl. Phys. {\bf A314}, 429 (1979).
\vspace{-8pt}
\bibitem{wong82}C. Toepffer and C. Y. Wong, Phys. Rev. C{\bf 25}, 
	1019 (1982).
\vspace{-8pt}
\bibitem{cas87}W. Cassing, Z. Phys. {\bf A465}, 317 (1987).
\vspace{-8pt}
\bibitem{abg94}P. Abgrall et al., Phys. Rev. C{\bf 49}, 1040 (1994).
\vspace{-8pt}
\bibitem{had96}F. Haddad et al., Z. Phys. {\bf A354}, 321 (1996).
\vspace{-8pt}
\bibitem{bor97}B. Borderie et al., Z. Phys. {\bf A357}, 7 (1997).
\vspace{-8pt}
\bibitem{bali95}B. A. Li and S. J. Yennello, Phys. Rev. C{\bf 52}, R1746 (1995).
\vspace{-8pt}
\bibitem{don97}R. Donangelo and S. R. Souza, Phys. Lett. {\bf B409}, 58 (1997).
\vspace{-8pt}
\bibitem{gatty75}B. Gatty et al., Z. Phys. {\bf A273}, 65 (1975).
\vspace{-8pt}
\bibitem{fed78}F. Beck, M. Dworzecka and H. Feldmeier, 
	Z. Phys. {\bf A289}, 113 (1978).
\vspace{-8pt}
\bibitem{udo84}W. U. Schr\"oder and J. R. Huizenga, in Treatise on Heavy-Ion 
	Science, ed. D. A. Bromley (Plenum, New York and London, 1984) 
	Vol. 2. P113.
\vspace{-8pt}
\bibitem{sherry1}S. J. Yennello et al., Phys. Lett. {\bf B321}, 15 (1994).
\vspace{-8pt}
\bibitem{sherry2}H. Johnston et al., Phys. Lett. {\bf B371}, 186 (1996).
\vspace{-8pt}
\bibitem{sherry3}H. Johnston et al., Phys. Rev. C{\bf 56}, 1972 (1997).
\vspace{-8pt}
\bibitem{sherry4}E. Ramakrishnan et al., Phys. Rev. C, to be published.
\vspace{-8pt}
\bibitem{bali97}B. A. Li, C. M. Ko and W. Bauer, International Journal of 
	Modern Phys. E, in press.
\vspace{-8pt}
\bibitem{bali96} B. A. Li, Z. Z. Ren, C. M. Ko and S. J. Yennello,
      Phys. Rev. Lett. {\bf 76}, 4492 (1996).
\vspace{-8pt}
\bibitem{pak1} R. Pak {\it et al.}, Phys. Rev. Lett. {\bf 78} 1022 (1997).
\vspace{-8pt}
\bibitem{pak2} R. Pak {\it et al.}, Phys. Rev. Lett. {\bf 78} 1026 (1997).
\vspace{-8pt}
\bibitem{lik97} B. A. Li, C. M. Ko and Z. Z. Ren,
      Phys. Rev. Lett. {\bf 78}, 1644 (1997).
\vspace{-8pt}
\bibitem{bauer92}W. Bauer, G. F. Bertsch and H. Schultz,
        Phys. Rev. Lett. {\bf 69}, 1888 (1992).
\vspace{-8pt}
\bibitem{ligross}B. A. Li and D. H. E. Gross, Nucl. Phys. {\bf A554}, 
257 (1993). 
 
\end{thebibliography}
\end{document}